\documentstyle[12pt,equation,oldlfont]{article}
\begin{document}
\newcommand{\tcr}{T_{cr}}
\newcommand{\ct}{\tilde{c}}
\newcommand{\df}{\delta \phi}
\newcommand{\dkl}{\delta \kappa_{\Lambda}}
\newcommand{\lx}{\lambda}
\newcommand{\Lx}{\Lambda}
\newcommand{\ex}{\epsilon}
\newcommand{\db}{{\bar{\delta}}}
\newcommand{\ks}{k_s}
\newcommand{\lb}{{\bar{\lambda}}}
\newcommand{\lbz}{\bar{\lambda}_0}
\newcommand{\lt}{\tilde{\lambda}}
\newcommand{\lr}{{\lambda}_R}
\newcommand{\lrt}{{\lambda}_R(T)}
\newcommand{\lbr}{{\bar{\lambda}}_R}
\newcommand{\lk}{{\lambda}(k)}
\newcommand{\lbk}{{\bar{\lambda}}(k)}
\newcommand{\lbkt}{{\bar{\lambda}}(k,T)}
\newcommand{\ltk}{\tilde{\lambda}(k)}
\newcommand{\mx}{{m}^2}
\newcommand{\mxk}{{m}^2(k)}
\newcommand{\mxkt}{{m}^2(k,T)}
\newcommand{\mb}{{\bar{m}}^2}
\newcommand{\mbu}{{\bar{\mu}}^2}
\newcommand{\mt}{\tilde{m}^2}
\newcommand{\mr}{{m}^2_R}
\newcommand{\mrt}{{m}^2_R(T)}
\newcommand{\mk}{{m}^2(k)}
\newcommand{\Pt}{\tilde P}
\newcommand{\Mt}{\tilde M}
\newcommand{\Qt}{\tilde Q}
\newcommand{\Nt}{\tilde N}
\newcommand{\rhb}{\bar{\rho}}
\newcommand{\rht}{\tilde{\rho}}
\newcommand{\rhz}{\rho_0}
\newcommand{\rhzt}{\rho_0(T)}
\newcommand{\yz}{y_0}
\newcommand{\rhzk}{\rho_0(k)}
\newcommand{\rhzkt}{\rho_0(k,T)}
\newcommand{\kx}{\kappa}
\newcommand{\kt}{\tilde{\kappa}}
\newcommand{\kk}{\kappa(k)}
\newcommand{\ktk}{\tilde{\kappa}(k)}
\newcommand{\Gammat}{\tilde{\Gamma}}
\newcommand{\Lt}{\tilde{L}}
\newcommand{\Zt}{\tilde{Z}}
\newcommand{\zt}{\tilde{z}}
\newcommand{\zh}{\hat{z}}
\newcommand{\uh}{\hat{u}}
\newcommand{\Mh}{\hat{M}}
\newcommand{\wt}{\tilde{w}}
\newcommand{\etat}{\tilde{\eta}}
\newcommand{\Gammak}{\Gamma_k}
\newcommand{\be}{\begin{equation}}
\newcommand{\ee}{\end{equation}}
\newcommand{\een}{\end{subequations}}
\newcommand{\ben}{\begin{subequations}}
\newcommand{\beq}{\begin{eqalignno}}
\newcommand{\eeq}{\end{eqalignno}}
\renewcommand{\thefootnote}{\fnsymbol{footnote} }
\pagestyle{empty}
\noindent
DESY 93-128 \\
HD-THEP-93-36 \\
September 1993
\vspace{3cm}
\begin{center}
{{ \Large  \bf
High Temperature Phase Transitions \\
without Infrared Divergences
}}\\
\vspace{10mm}
N. Tetradis \\
{\em Deutsches Elektronen-Synchrotron DESY, Gruppe Theorie,\\
Notkestr. 85, 22603 Hamburg}\\
\vspace{5mm}
and \\
\vspace{5mm}
C. Wetterich \\
{\em
Institut f\"ur Theoretische Physik, Universit\"at Heidelberg,\\
Philosophenweg 16, 69120 Heidelberg}
\end{center}

\setlength{\baselineskip}{20pt}
\setlength{\textwidth}{13cm}

\vspace{3.cm}
\begin{abstract}
{
The most commonly used method for the study of high temperature phase
transitions is based on the perturbative evaluation of the temperature
dependent effective potential. This method becomes unreliable in the
case of a second order or weakly first order phase transition, due to
the appearance of infrared divergences. These divergences can be
controlled through the method of the effective average action which
employs renormalization group ideas. We report on the study of the
high temperature phase transition for the $N$-component $\phi^4$ theory.
A detailed quantitative picture of the second order phase transition
is presented, including the critical exponents for the behaviour
in the vicinity of the critical temperature.
An independent check of the results is obtained in the large
$N$ limit, and contact with the perturbative approach is established through
the study of the Schwinger-Dyson equations.
}
\end{abstract}
\clearpage
\setlength{\baselineskip}{15pt}
\setlength{\textwidth}{16cm}
\pagestyle{plain}
\setcounter{page}{1}

\newpage

\setcounter{equation}{0}
\renewcommand{\theequation}{{\bf 1.}\arabic{equation}}

\section*{1.
Introduction}

Spontaneous symmetry breaking is one of the most
prominent features of the standard model of electroweak interactions.
The masses of the gauge bosons and fermions are proportional to the
vacuum expectation value of the Higgs doublet.
According to the original argument of Kirzhnits and Linde \cite{orig}
this expectation value vanishes at sufficiently high temperature
and the electroweak symmetry is restored.
Such high temperatures were presumably realized in the very early
universe immediately after the big bang. As the universe
cooled there must have been a phase transition from the
symmetric
to the spontaneously broken phase of the standard model.
This phase transition may have many important consequences for our
present universe,
one example being the possible creation of the excess of matter
compared to antimatter (baryon asymmetry)
\cite{shaposh}.
The physical implications of the high temperature electroweak phase transition
depend strongly
on its nature (whether it is second or first order) and its details.
More specifically, the possibility of generating the baryon asymmetry
imposes the requirement of out of equilibrium conditions
\cite{sakharov}, which is
satisfied only if the phase transition is of the first order;
the exact amount of produced baryon number is very sensitive to the
details of the fluctuations which drive the transition (profile
of bubbles, velocity of the wall etc.) \cite{models}; and
avoiding the washing out of any generated baryon asymmetry requires a
sufficiently strong first order phase transition \cite{bound}.
\par
None of the above questions has a satisfactory answer in the framework
of the perturbative evaluation of the
effective action
\cite{colwein} and its extension to non-zero
temperature \cite{linde,doljack,weinberg} which is the usual tool for
such investigations. The reason is the appearance of infrared divergences
which result from high order
contributions in the perturbative expansion. More specifically for the
evaluation of the effective potential,
higher loop corrections involving scalar fields give contributions
proportional to arbitrary powers of
$\frac{\lx T}{m_s(T)}$, which diverge near the critical temperature
where $m_s(T_{cr})=0$ \cite{linde,doljack,weinberg}.
If Goldstone modes are present the situation is more difficult, since their
masses vanish at all temperatures in the spontaneously broken phase.
Similarly loops with gauge fields involve powers
of $\frac{g^2 T}{m_A(T)}$, which are problematic in the whole symmetric phase
\cite{gaugediv}. It is interesting to observe that the combinations
$\lx T$, $g^2 T$ can be viewed as effective three-dimensional couplings
resulting from the compactification of one dimension
in the non-zero temperature formalism.
It becomes apparent then, that the infrared divergences are of
the same nature as the infrared
divergences of three-dimensional field theories.
The formalism which can control
the infrared regime of non-zero temperature
four-dimensional theories must provide a reliable treatment of
the infrared singularities for field theories
in dimensions less than four.
\par
In order to reliably compute the long distance behaviour we
follow Wilson's philosophy \cite{wilson} of
iteratively integrating out the short distance
modes and using an effective action to describe their physical effects.
In the original proposal an ultraviolet cutoff $\Lambda$ was introduced
and the change of the ``cutoff action'' $S_{\Lambda}$ as a function of
$\Lambda$ was described by an ``exact renormalizaton group equation''
\cite{wilson,wegner,polchinski}.
Lowering $\Lambda$ to zero should, in principle,
express the long distance properties of the theory in terms of
microscopic properties defined for large $\Lambda$. Unfortunately, the
``cutoff action'' $S_{\Lambda}$ does not become the
generating functional of the 1PI Green functions in the limit
$\Lambda \rightarrow 0$, but rather a relatively complicated functional
related to the connected Green functions, with sources expressed in terms
of fields \cite{keller}. Correspondingly, the form of the exact
renormalization group equation does not exhibit the infrared properties
of the theory in an obvious way. So far it has been only possible to
use the exact renormalization group equation in the context
of perturbation theory (where the usual infrared problems are present)
or the $1/N$ expansion \cite{ellver}.
\par
We follow here a new approach \cite{christof1,christof2,exact}
based on the intuitive picture of an effective action for averages of
fields, formulated in continuous space. The average is taken over a
volume of size $k^{-d}$ and the corresponding effective
average action $\Gamma_k$ depends on the scale $k$. The new evolution
equation \cite{exact} which describes the dependence of $\Gamma_k$ on
$k$ is manifestly governed by the infrared properties of the theory.
With the help of this equation the infrared problems can be resolved
even for two- and three-dimensional theories in a non-perturbative
context \cite{christof2}.
The evolution equation for $\Gamma_k$ is an exact non-perturbative
equation \cite{exact}.
Its solution interpolates between the classical action for large $k$
and the effective action for $k \rightarrow 0$.
It has been shown that for $k=0$ the effective average action is the
generating functional for the 1PI Green functions \cite{exact}.
All relevant properties of the ground state, masses and interactions
can, therefore, be extracted from the solution for
$\Gamma_k$ in a straighforward manner. Furthermore, for $k > 0$ the
effective average action describes a ``coarse grained effective action''
in the sense used by Langer \cite{langer} in his treatment of
first order phase transitions. Since for $k > 0$ only the quantum fluctuations
with momenta $q^2 \geq k^2$ have been incorporated, the effective average
action $\Gamma_k$ is the appropriate functional for all processes with an
effective physical infrared cutoff of the order of $k$. In short, the effective
average action is the conventional effective action
with an additional explicit infrared cutoff suppressing
the contributions to the functional integral coming from quantum fluctuations
with small momentum $q^2 < k^2$.
All the internal symmetries, including gauge and chiral symmetries
in the case of gauge fields or chiral fermions, can be maintained by the
infrared cutoff. Our
formulation is carried out in continuous space, thus preserving the
symmetries of rotations and translations.
This is important, since it
permits a simple and efficient parametrization of
$\Gamma_k$.
The presence of a separate cutoff (other than masses, field
expectation values or momenta) guarantees that no infrared divergence can
appear for $k > 0$.
Moreover, the limit
$k \rightarrow 0$ can be taken in a controlled way so that no singularities
are encountered.
\par
The practical use of formally exact evolution equations depends primarily
on the existence of viable approximation schemes since systems of differential
equations for infinitely many variables are usually not exactly
solvable
\footnote{A few exceptions exist for limiting cases, such as
$N \rightarrow \infty$ where the system of equations becomes decoupled.}.
One needs an efficient parametrization of $\Gamma_k$ which allows the
reduction of the exact evolution equation to a few equations for running
couplings. This parametrization is model dependent and involves physical
intuition. One needs essentially a guess about the relevant degrees
of freedom and the general form of their propagators. In this review
we concentrate
on the $O(N)$-symmetric $\phi^4$ theory \cite{transition, largen, indices},
which has served for years as
the prototype for investigations concerning the question of symmetry
restoration
at high temperature.
(More complicated scalar theories, as well as gauge theories,
can be approached along
similar lines and such investigations are under way
\footnote{For applications of the renormalization group to the study of
first order phase transitions also see ref. \cite{amit} and references therein.
For a recent study see ref. \cite{march}.}.)
The physical importance of the theory
arises from the fact that, for $N=4$, it describes the scalar
sector of the electroweak standard model
in the limit of vanishing gauge and Yukawa couplings, and
is relevant for the QCD phase transition
for vanishing up and down quark mass \cite{wilczek}.
In the following section we give a detailed definition of the
effective average action and present the exact evolution equation which
permits its determination from the classical action.
We also develop a parametrization scheme
which allows for the determination
of physical quantities such as masses, couplings,
wave function renormalizations etc.
In section 3 we generalize the formalism in order to take into account
non-zero temperature effects and discuss the phase transition for the
$O(N)$-symmetric $\phi^4$ theory.
In section 4
we study, with various techniques, the phase transition in the large
$N$ limit.
This allows us to
obtain an independent check of the results of section 3
and make contact with the standard perturbative treatments.
Finally, in section 5 we give a summary of various other works
employing the effective average action and comment on the
application of the scheme to the study of tunneling phenomena and
first order phase transitions.
We conclude by discussing briefly the extension of the
method to the case of more complicate scalar as well as gauge theories.

\setcounter{equation}{0}
\renewcommand{\theequation}{{\bf 2.}\arabic{equation}}

\section*{2. The effective average action}

We consider a theory of $N$ real scalar fields $\chi^a$, in
$d$ dimensions, with an
$O(N)$-symmetric action $S[\chi]$.
We specify the action together with some ultraviolet
cutoff $\Lambda$, so that the theory is properly
regulated.
We add to the kinetic term an infrared regulating piece \cite{exact}
\be
\Delta S = \frac{1}{2} \int d^d q
R_k(q) \chi^*_a(q) \chi^a(q),
\label{twoone} \ee
where $\chi^a(q)$ are the Fourier modes of the scalar fields.
The function $R_k$  is employed in
order to prevent the propagation of modes
with characteristic momenta $q^2 < k^2$.
This can be achieved, for example,
by the choice
\be
R_k(q) = \frac{Z_k q^2 f^2_k(q)}{1 - f^2_k(q)},
\label{twotwo} \ee
with
\be
f_k(q) = \exp \biggl\{ - a \left( \frac{q^2}{k^2} \right)^b
\biggr\},
\label{twothree} \ee
where $a$, $b$ are
constants of order one which determine the shape of the cutoff.
As a result the inverse propagator
derived from the action $S + \Delta S$ has a minimum $\sim k^2$.
The modes with $q^2 \gg k^2$ are unaffected,
while the
low frequency modes with $q^2 \ll k^2$ are cut off.
The quantity $Z_k$ in eq. (\ref{twotwo})
is an appropriate wave function renormalization whose
precise definition will be given below.
We emphasize at this point that the above form of $R_k$ is not unique
and many alternative choices are possible.
The physical quantities
which are obtained when the cutoff is removed ($k \rightarrow 0$)
are independent of the form of $R_k$.
For the choice of eq. (\ref{twothree}) the values of $a$ and
$b$ should not affect the physical results and can be chosen freely.
For a discussion of a natural choice of
$a$, $b$ and the residual $a$ and $b$ dependence when approximations
are used we refer the reader to refs. \cite{exact, transition}.
We subsequently introduce sources and
define the generating functional for the connected Green functions
for the action $S + \Delta S$. Through a Legendre
transformation we obtain the
generating functional for the 1PI Green functions
${\tilde \Gamma}_k[\phi^a]$, where $\phi^a$ is the expectation value of the
field $\chi^a$ in the presence of sources.
The use of the modified propagator for the calculation of
${\tilde \Gamma}_k$ results in the effective integration of only the
fluctuations with $q^2 \geq
k^2$. Finally, the effective average action is
obtained by removing the infrared cutoff
\be
\Gamma_k[\phi^a] = {\tilde \Gamma}_k[\phi^a] -
\frac{1}{2} \int d^d q
R_k(q) \phi^*_{a}(q) \phi^a(q).
\label{twofour} \ee
For $k$ equal to the ultraviolet cutoff $\Lambda$, $\Gammak$ becomes
equal
to the classical action $S$ (no effective integration of modes takes
place), while for $k \rightarrow 0$ it tends towards the effective action
$\Gamma$ (all the modes are included)
which is the generating functional of the 1PI Green functions computed
from $S$ (without infrared cutoff).
The interpolation of $\Gammak$ between the classical and the
effective action makes it a very useful field theoretical tool.
The means for practical calculations is provided by an exact
evolution equation which describes the response of the
effective average action to variations of the infrared cutoff
($t=\ln (k/\Lambda)$) \cite{exact}
\be
\frac{\partial}{\partial t} \Gammak[\phi]
= \frac{1}{2} {\rm Tr} \bigl\{ (\Gammak^{(2)}[\phi] + R_k)^{-1}
\frac{\partial}{\partial t} R_k \bigr\}.
 \label{twofive} \ee
Here $\Gammak^{(2)}$ is the second functional derivative of the effective
average action with respect to $\phi^a$. For
real fields it reads in momentum space
\footnote{
In order to avoid excessive formalism, we shall use the same notation
for functions or operators in position space and their Fourier transforms,
defined with the convention
$\phi(q)=(2 \pi)^{-\frac{d}{2}}\int d^dx \exp(iq_{\mu}x^{\mu}) \phi(x)$.
}
\be
(\Gammak^{(2)})^a_b(q,q') =
\frac{\delta^2 \Gammak}{\delta \phi^*_a(q) \delta \phi^b(q')},
\label{twosix} \ee
with
\be
\phi^a(-q)=\phi^*_a(q).
\label{twoseven} \ee
The exact evolution equation gives the response of the
effective average
action $\Gammak$ to variations of the scale $k$, through a one-loop expression
involving the exact inverse propagator $\Gamma^{(2)}_k$ together with an
infrared cutoff provided by $R_k$.
It has a simple graphical representation (fig. 1). Our non-perturbative
exact evolution equation can be
viewed as a partial differential equation for
the infinitely many variables $t$ and $\phi^a(q)$.
Its usefulness depends on the existence of
appropriate truncations which permit its solution.
\par
Before presenting the formalism which leads to approximate solutions of
eq. (\ref{twofive}), we briefly comment on the relation of the effective
average action to the average action, which
was introduced in ref. \cite{christof1,christof2}
in order to describe the dynamics of averages of fields over volumes
$\sim k^{-d}$.
The field average is the realization in the continumm
of the block spin idea of Kananoff and Wilson \cite{kadanoff, wilson}.
The advantage of the average action
over the block spin action lies in the preservation of the symmetries
of rotations and translations, which
permit an efficient parametrization.
In the context of the average action the infrared cutoff $k$ is
naturally introduced through
the averaging procedure. With an appropriate implementation of the averaging
procedure \cite{more} the average action
is identical to the effective average action
up to an explicitely known regulator term.
The dependence of the average action on $k$
is computed through
an one-loop renormalization group equation which corresponds to
a truncation of
eq. (\ref{twofive}).
Moreover, within the approximations used so far there
is no difference for low
momentum quantities between the average action
\cite{christof1,christof2,more} and
the effective average action \cite{exact}.
As a result, all previous calculations of the average action
can be viewed as
approximate solutions of eq. (\ref{twofive}) with some appropriate truncation.
We finally should mention that the formal relation between our exact evolution
equation \cite{exact} and the ``exact renormalization group equation''
\cite{wilson,wegner,polchinski}
has by now been established \cite{priv}.
\par
As we have already mentioned, for the solution of eq. (\ref{twofive})
one has to develop an efficient truncation scheme.
We consider
an effective average action of the form
\be
\Gammak =
\int d^dx \bigl\{ U_k(\rho)
+ \frac{1}{2} \partial^{\mu} \phi_a Z_k(\rho)
\partial_{\mu} \phi^a
+ \frac{1}{4} \partial^{\mu} \rho Y_k(\rho)
\partial_{\mu} \rho
\bigr\},
\label{twoeight} \ee
where $\rho=\frac{1}{2} \phi_a \phi^a$.
In order to turn the evolution equation for the
effective average action into equations for
$U_k$, $Z_k$ and $Y_k$, we have to evaluate the trace in eq. (\ref{twofive})
for properly chosen
background field configurations.
For the evolution equation for $U_k$ we have to expand around a constant
field configuration (so that the derivative terms in the parametrization
(\ref{twoeight}) do not contribute to the l.h.s. of eq. (\ref{twofive})).
Eq. (\ref{twofive}) then gives \cite{christof2,exact,indices}
\be
\frac{\partial}{\partial t} U_k(\rho) =
\frac{1}{2} \int \frac{d^d q}{(2 \pi)^d}
\left( \frac{N-1}{M_0} +\frac{1}{M_1} \right)
\frac{\partial}{\partial t} R_k(q),
\label{twonine} \ee
with $R_k(q)$ given by eqs. (\ref{twotwo}), (\ref{twothree}),
\beq
M_0(\rho) =~&Z_k(\rho) q^2 + R_k(q) + U'_k(\rho) \nonumber \\
M_1(\rho) =~&{\tilde Z}_k(\rho)  q^2
+ R_k(q) + U'_k(\rho) + 2 \rho U''_k(\rho)
\label{twoten} \eeq
and
\be
{\tilde Z}_k(\rho)  = Z_k(\rho) + \rho Y_k(\rho).
\label{twoeleven} \ee
Here primes denote derivatives with respect to $\rho$.
For simplicity we have considered only the $\rho$ dependence of
$Z_k$ and $\Zt_k$. If their dependence on $q^2$ is also included
the evolution equation (\ref{twonine}) becomes exact.
For $\rho$ different from zero it is easy to recognize the
first term in eq. (\ref{twonine}) as the contribution from the
$N-1$ Goldstone bosons ($U'$ vanishes at the minimum).
The second term is then related to the radial mode.
The evolution of the wave function renormalizations
$Z_k$, $\Zt_k$, which leads to the
determination of the anomalous dimensions, requires an expansion around
a background with a small momentum dependence.
We do not give the explicit results of this calculation here, but we
refer the reader to refs. \cite{christof2, indices} for the details.
The evolution equations for $U_k$, $Z_k$ and $Y_k$ are partial differential
equations for independent variables $t$ and $\rho$. In most cases
their study is
possible when they are
turned into an (infinite) set of coupled ordinary differential equations for
independent variable $t$.
This is achieved by Taylor expanding
$U_k$, $Z_k$ and $Y_k$ around some value
of $\rho$. Since we are interested in the vacuum structure of the
theory, we can use an expansion around
the $k$ dependent minimum $\rhzk$ of $U_k$.
In the limit $k \rightarrow 0$ this minimum
specifies the vacuum of the theory, and
$U=U_0$, $Z=Z_0$ and $Y=Y_0$
and their $\rho$-derivatives at the minimum give the
renormalized masses, couplings and wave function renormalizations.
We can now define the quantity $Z_k$ appearing in
eq. (\ref{twotwo})
as
\be
Z_k = Z_k(\rhzk).
\label{twotwelve} \ee
For studies which
concentrate on the minimum of $U_k$,
the above definition permits the combination of
the leading kinetic contribution to
$\Gammak^{(2)}$ and $R_k$ in eq. (\ref{twosix}), into an effective
inverse propagator (for massless fields)
\be
Z_k P(q^2) = Z_k q^2 + R_k = \frac{Z_k q^2}{1 - f^2_k(q)},
\label{twothirteen} \ee
with $f_k(q)$ given by eq. (\ref{twothree}).
For $q^2 \gg k^2$ the inverse ``average'' propagator $Z_k P(q)$ approaches
the standard inverse propagator $Z_k q^2$ exponentially fast, whereas
for $q^2 \ll k^2$ the infrared cutoff
prevents the propagation.
\par
In this section
we have summarized the basic formalism of the effective average action
at zero temperature. In the
following we generalize it for the case of non-zero temperature and
use appropriate truncations in order to study
the high temperature phase transition for the $O(N)$-symmetric
$\phi^4$ theory.

\setcounter{equation}{0}
\renewcommand{\theequation}{{\bf 3.}\arabic{equation}}

\section*{3.
The high temperature phase transition for the $O(N)$-symmetric
$\phi^4$ theory}

{\bf 3a) Approximations and truncations:}
A first approximation, which results in considerable simplification
in the formalism, is the omission of the effects
of wave function renormalization.
This is expected to be a good approximation
since we shall be dealing with an effective three-dimensional
critical theory
for which the anomalous dimension is known to be $\eta \simeq 0 - 0.04$.
At the end of the section we shall relax this approximation in order
to perform an accurate calculation of the critical exponents for the
three-dimensional theory.
For the time being we
concentrate on eq. (\ref{twonine}) in which we set $Z_k=1$, $Y_k=0$,
$\Zt_k=1$. We can, therefore, define an effective inverse
propagator $P$ according to
eq. (\ref{twothirteen}) (with $Z_k=1$)
and rewrite eq. (\ref{twonine}) as
(($t=\ln (k/\Lambda)$,
$x = q^2$)
\beq
\frac{\partial}{\partial t} U_k(\rho)
= & \frac{1}{2} (2 \pi)^{-d} \int d^d q
\frac{\partial P}{\partial t}
\left( \frac{1}{P+U'_k(\rho)+2U''_k(\rho)\rho} + \frac{N-1}{P+U'_k(\rho)}
\right) \nonumber \\
= & v_d \int_0^{\infty}dx x^{\frac{d}{2}-1}
\frac{\partial P}{\partial t}
\left( \frac{1}{P+U'_k(\rho)+2U''_k(\rho)\rho} + \frac{N-1}{P+U'_k(\rho)}
\right),
\label{threeone} \eeq
with
\be
v_d^{-1} = 2^{d+1} \pi^{\frac{d}{2}} \Gamma\left(\frac{d}{2}\right).
\label{threetwo} \ee
We remind the reader that primes denote derivatives with respect to
$\rho$. Equation (\ref{threeone}) is a partial differential equation
for independent variables
$t$ and $\rho$, which must be solved with the boundary condition
$U_\Lambda(\rho) = V(\rho)$ ($V(\rho)$ being the classical potential).
Its solution for $k \rightarrow 0$
gives the effective potential $U(\rho) = U_0(\rho)$.
We are interested in solving (\ref{threeone})
around the minimum of the effective average potential.
In order to do so we parametrize $U_k$ in terms of the location of its
minimum and its successive derivatives with respect to $\rho$ at the minimum.
This results in an infinite system of coupled differential equations, which
we solve approximately
by truncating at a finite number of equations and keeping an
equal number of $\rho$-derivatives. As a first step we truncate after
the second derivative of the potential.
\par
In the {\it spontaneously broken regime}
the effective
average potential $U_k$ has its minimum at $\rho_0(k) \not= 0$, where
$ U'_k(\rho_0) = 0. $ We define
\be
\lb(k) = U''_k(\rhzk),
\label{threethree}
\ee
and obtain the following
evolution equations for the scale dependence of $\rho_0(k)$ and $\lb(k)$
(for the details of the derivation see refs.
\cite{christof2,transition,indices})
\beq
\frac{d \rhz}{dt} = &
-v_d k^{d-2} \bigl\{ 3 L^d_1(2 \lb \rhz) + (N-1) L^d_1(0) \bigr\}
\label{threefour} \\
\frac{d \lb}{dt} = &
-v_d k^{d-4} \lb^2 \bigl\{ 9 L^d_2(2 \lb \rhz) +
(N-1) L^d_2(0) \bigr\},
\label{threefive}
\eeq
with the dimensionless integrals $L^d_n(w)$ given by
\beq
L^d_n(w) = &
- n k^{2n-d} \pi^{-\frac{d}{2}} \Gamma \left( \frac{d}{2} \right)
\int d^d q \frac{\partial P}{ \partial t} (P + w)^{-(n+1)} \nonumber \\
= &
- n k^{2n-d}
\int_0^{\infty} dx x^{\frac{d}{2}-1}
\frac{\partial P}{ \partial t} (P + w)^{-(n+1)},
\label{threesix} \eeq
and $P$ given by eq. (\ref{twothirteen}) (with $Z_k=1$). \\
In the {\it symmetric regime} ($\rhzk=0$) we define
\beq
\mx = & U'_k(0)
\label{threeseven} \\
\lbk = & U''_k(0),
\label{threeeight} \eeq
and obtain the following evolution equations
\beq
\frac{d \mx}{d t} = &
{}~(N+2) v_d k^{d-2} \lb L^d_1(\mx)  \label{threenine} \\
\frac{d \lb}{d t} = &
- (N+8) v_d k^{d-4} \lb^2 L^d_2(\mx).  \label{threeten}
\eeq
The systems of equations (\ref{threefour}),(\ref{threefive}) and
(\ref{threenine}),(\ref{threeten})
can be solved for given ``short distance values"
$\rhz(k=\Lambda)$ and $\lb(k=\Lambda)$.
We start at the cutoff $\Lambda$ and follow the renormalization group
flow towards the infrared $(k \rightarrow 0)$.
In this way we obtain the ground state of the theory $\rhz = \rhz(k=0)$,
as well as the renormalized couplings, mass terms etc.
It should be noted that, even when one starts in the broken regime at
$k=\Lambda$, it is possible that the evolution,
as given by the first set of differential equations, may drive $\rhzk$ to
zero at some non-zero $k_s$. From that point on
the theory is in the symmetric regime
and one has to continue the evolution using the second set of
equations, with boundary conditions $\mx (k_s)=0$ and
$\lb (k_s)$ given by its value at the end of the running in the broken regime.
\par
The integrals $L^d_n(w)$, defined in eq. (\ref{threesix}),
have been discussed extensively in refs.
\cite{christof2,indices,convex} for various shapes of the
cutoff, as determined by the
values of the parameters $a$ and $b$ in eq. (\ref{twothree}).
As an example, we plot in fig. 2 the integrals $L^4_1(w)$,  $L^4_2(w)$.
Their most interesting property, for our discussion, is that
they fall off for large values of $w/k^2$, following a power law. As
a result they introduce threshold behaviour for the
contributions of massive modes to the evolution equations.
When the ``running'' squared
mass $2 \lb \rhz$ or $\mx$ becomes much larger than the scale
$k^2$ (at which the system is probed and the fields are averaged), these
contributions vanish and the massive modes decouple.
We evaluate the integrals $L^d_n(w)$
numerically and use numerical fits for the solution of
the evolution equations.
\par
{\bf 3b) The non-zero temperature formalism:}
In order to extend our expressions to the non-zero temperature case
we only need to recall that, in Euclidean formalism, non-zero temperature $T$
results
in periodic boundary conditions in the time direction (for bosonic fields),
with periodicity $1/T$ \cite{kapusta}.
This leads to a discrete spectrum for the zero component of the momentum
$q_0$
\be
q_0 \rightarrow 2 \pi m T,~~~~~~~~~m=0,\pm1,\pm2,...
\label{threeeleven} \ee
As a consequence the integration over $q_0$ is replaced by summation over the
discrete spectrum
\be
\int \frac{d^d q}{(2 \pi)^d} \rightarrow
T \sum_m \int \frac{d^{d-1}\vec{q}}{(2 \pi)^{d-1}}.
\label{threetwelve} \ee
With the above remarks in mind we can easily generalize equation
(\ref{threeone}) in order to take into account the temperature effects.
For the temperature dependent effective average potential $U_k(\rho,T)$
we obtain:
\be
\frac{\partial}{\partial t} U_k(\rho,T)
= \frac{1}{2} (2 \pi)^{-(d-1)} T \sum_n \int d^{d-1} \vec{q}~~
\frac{\partial P}{\partial t}
\left( \frac{1}{P+U'_k(\rho,T)+2U''_k(\rho,T)\rho} + \frac{N-1}{P+U'_k(\rho,T)}
\right),
\label{threethirteen} \ee
with the implicit replacement
\be
q^2 \rightarrow \vec{q}^2 + 4 \pi^2 m^2 T^2
\label{threefourteen} \ee
in $P$.
Again, the usual temperature dependent effective potential
\cite{linde, doljack, weinberg}
obtains from $U_k(\rho,T)$ in the limit $k \rightarrow 0$.
As before, we can parametrize $U_k(\rho,T)$ in terms of its minimum and
its derivatives at the minimum. The evolution equations are given by
(\ref{threefour}),(\ref{threefive}) and
(\ref{threenine}),(\ref{threeten})
with the obvious generalizations
$ \rhzk  \rightarrow  \rhzkt$ etc.
\par
The momentum integrals for non-vanishing temperature read:
\be
L^d_n(w,T) =
- n k^{2n-d} 2 \pi^{-\frac{d}{2}+1} \Gamma \left( \frac{d}{2} \right)
 T \sum_m
\int d^{d-1} \vec{q}~~ \frac{\partial P}{ \partial t} (P + w)^{-(n+1)},
\label{threesixteen} \ee
where the implicit replacement
(\ref{threefourteen}) is again assumed in $P$.
Their basic properties can be established
analytically.
For $T \ll k$ (low temperature region) the summation over discrete
values of $m$ in expression (\ref{threesixteen}) is equal to the
integration over a continuous range of $q_0$ up to exponentially small
corrections.
Therefore
\be
L^d_n(w,T) = L^d_n(w)~~~~~~~~~{\rm for}~~T \ll k.
\label{threeseventeen} \ee
In the opposite limit $T \gg k$ (high temperature region)
the summation over $m$ is
dominated by the $m=0$ contribution.
Terms with non-zero values of $m$ are suppressed
by $  \sim \exp \left( -( mT/k)^{2 b} \right).$
The leading contribution gives the
the simple expression
\be
L^d_n(w,T) = \frac{v_{d-1}}{v_d} \frac{T}{k} L^{d-1}_n(w)~~~~~~{\rm
for}~~T \gg k,
\label{threeeighteen} \ee
with $v_d$ defined in (\ref{threetwo}).
The two regions of $T/k$ in which $L^d_n(w,T)$ is given by the
equations (\ref{threeseventeen}) and (\ref{threeeighteen})
are connected by a small interval in which
the exponential corrections result in complicated dependence on
$w$ and $T$ (intermediate region).
The above conclusions are verified by a numerical calculation of
$L^4_1(w,T)$ and $L^4_2(w,T)$.
In fig. 3 we plot $L^4_1(w,T)/L^4_1(w)$ as a function of $T/k$,
for various values of $w/k^2$ and for $b=3$, $a = 0.952$.
The three regions (low temperature, intermediate and high temperature)
are separated by vertical lines.
\par
{\bf 3c) Running in four and three dimensions:}
{}From this point on we concentrate on the four-dimensional theory.
For zero temperature the relevant evolution equations in the
spontaneously broken regime are given
by (\ref{threefour}),(\ref{threefive})
with $d=4$. Eq. (\ref{threefour}) results in the ``quadratic renormalization''
of the minimum of the effective average potential \cite{christof1, christof2}:
Apart from threshold effects, $\rhzk$ evolves proportionally to $k^2$. This
is the reflection in our language of the quadratic divergences of standard
perturbation theory. This behaviour is
depicted by the solid line of fig. 4 for $N=4$.
In the limit $k \rightarrow 0$, $\rhzk$ settles down at a value which
corresponds to the renormalized minimum of the effective potential
(the zero temperature vacuum of the theory) which we denote by
\be
\rhz = \rhz (0).
\label{threenineteen} \ee
Eq. (\ref{threefive}) results in the usual
logarithmic running for the quartic coupling in four dimensions.
The presence of the
integrals $L^4_n$ introduces threshold effects in the evolution. As
a result the logarithmic running of the coupling stops as $k \rightarrow 0$
for $N=1$, due to the decoupling of the massive radial mode. In the
case $N \not= 1$, due to the presence of massless Goldstone modes,
the logarithmic running never stops and this results in
an infrared free theory. The running of $\lbk$ is given by the solid line
of fig. 5 for $N=4$. Since $\lb(k=0) = 0$ for
$N \not= 1$, we define the renormalized
coupling at some non-zero $k$ according to
\be
\lr = \lb \left( \sqrt{2 \lr \rhz} \right)
\label{threetwenty} \ee
for all $N$.
\par
For non-zero temperature we have to distinguish between the three
different regions as $k$ decreases: \\
I. In the low temperature region ($k \gg T$) the evolution is purely
four-dimensional. \\
II. In the intermediate region the running is complicated and we can
study it only numerically. \\
III. In the high temperature region ($k \ll T$) we can use
eq. (\ref{threeeighteen}) in order to rewrite the evolution equations in
the spontaneously broken regime as
\beq
\frac{d \rho'_0}{dt} = &
-v_3 k \bigl\{ 3 L^3_1 ( 2 \lb' \rho'_0 )
+ (N-1) L^3_1(0) \bigr\}
\label{threetwentyone} \\
\frac{d \lb'}{dt} = &
-v_3 k^{-1} (\lb')^2
\bigl\{ 9 L^3_2 ( 2 \lb' \rho'_0 )
+ (N-1) L^3_2(0) \bigr\},
\label{threetwentytwo} \eeq
with
\beq
\rho'_0(k,T) = & \frac{\rhzkt}{T} \nonumber \\
\lb'(k,T) =& \lbkt T.
\label{threetwentythree} \eeq
Comparison with
eqs.
(\ref{threefour}), (\ref{threefive})
shows that the above equations are exactly the
ones of the three-dimensional theory for the effective three-dimensional
couplings $\rho'_0(k,T)$, $\lb'(k,T)$. We recover
the fact that the behaviour of the high temperature four-dimensional
theory is determined by the zero temperature three-dimensional theory.
Moreover, we have developed a formalism which connects the four-dimensional
regime with the effective three-dimensional one in a quantitative way.
This is crucial since the precise initial values of the
three-dimensional running are needed for the determination of the
critical temperature etc.
\par
It is convenient again to define the dimensionless quantities
\beq
{\kx}(k,T) = &~\frac{\rhz'(k,T)}{k} = \frac{\rhzkt}{k T} \nonumber \\
{\lx}(k,T) = &~\frac{\lb'(k,T)}{k} = \lbkt \frac{T}{k}.
\label{threetwentyfour} \eeq
In terms of these quantities the evolution equations read:
\beq
\frac{d \kx}{dt} = & - \kx
-v_3 \bigl\{ 3 L^3_1 ( 2 \lx \kx )
+ (N-1) L^3_1(0) \bigr\}
\label{threetwentyfive} \\
\frac{d \lx}{dt} = &~- \lx
-v_3 \lx^2
\bigl\{ 9 L^3_2 ( 2 \lx \kx )
+ (N-1) L^3_2(0) \bigr\}.
\label{threetwentysix}
\eeq
The main characteristic of the last equations
arises from
the term $-\lx$ on the r.h.s. of (\ref{threetwentysix}), which is
due to the dimensions of $\lb'$. In consequence, the
dimensionless quartic coupling $\lx$ is not infrared free. Its
behaviour with $k \rightarrow 0$ is characterized by an
approximate fixed point for the region where $\kx$ varies only
slowly. Taken together, the
pair of differential equations
for $(\kx, \lx)$
has an exact fixed point
($\kx_{fp}$, $\lx_{fp}$) corresponding to
the phase transition \cite{christof2, transition}.
This can be demonstrated explicitly by the numerical solution of
(\ref{threetwentyfive}), (\ref{threetwentysix}).
The phase diagram for $N=4$ is
plotted in fig. 6.
There is a critical line separating the spontaneously broken from
the symmetric phase. We also have indicated the flow of the couplings for
decreasing $k$. As $k$ is lowered the system enters the high temperature
region at some point in the phase diagram.
When this point is very close to the
critical line the system spends very long ``time'' $|t|=-\ln(k/\Lambda)$
near the
critical point, before deviating either towards the spontaneously
broken or the symmetric phase. During this ``time'' it loses memory of the
initial conditions.
As a result the dynamics near the phase transition is determined by
the fixed point, independently of the ``short distance'' values of the
couplings. Moreover, the character of the dynamics is
purely three-dimensional.
\par
The solution of the evolution equations for increasing temperature is
plotted in figs. 4 and 5 for $N=4$.
All the solutions correspond to the same ``short distance''
theory as specified by the same initial
conditions at $k = \Lambda \gg T$.
In fig. 4 we observe
the deviation from the zero temperature behaviour which starts
with the complicated running in the
intermediate region.
For low temperatures, in the limit $k \rightarrow 0$, $\rhzkt$ reaches
an asymptotic value $\rho_0(0,T) < \rhz$.
This value corresponds to the vacuum expectation value
of the non-zero temperature
theory and we denote it by
\be
\rhzt=\rhz(0,T).
\label{threetwentyseven} \ee
At a specific temperature $T_{cr}$, $\rhzt$ becomes zero and this
signals the restoration of symmetry for $T \geq T_{cr}$.
In fig. 5, $\lbkt$ deviates from the
zero temperature running and, for $T \le T_{cr}$,
approaches zero $\sim k$.
The latter effect is due to the fluctuations of the massless
Goldstone bosons. For $N=1$, as
in the zero temperature case, the running of the
coupling is stopped when the massive radial mode decouples.
In the vicinity of the critical temperature, however,
the radial mode remains massless for most of the evolution. As a result
the coupling runs to zero $\sim k$.
For $N \not= 1$
we see that the problem of the definition of the renormalized
coupling reappears for the non-zero temperature theory in the
spontaneously broken phase for $N \not= 1$.
In analogy to (\ref{threetwenty}) we define $\lrt$ at a non-zero scale
\be
\lrt = \lb \bigl( \sqrt{2 \lrt \rhzt}, T \bigr)
\label{threetwentyeight} \ee
in the spontaneously broken phase for all values of $N$.
For $T > T_{cr}$ the running in the spontaneously broken regime ends
at a non-zero $\ks$,
at which $\rhz(\ks,T)$ equals zero.
{}From this point on we continue the evolution in the symmetric regime.
The running of $\mxkt$ is depicted in fig. 4 while the evolution of
$\lbkt$ proceeds continuously in the new
regime
as shown in fig. 5.
In the symmetric phase the theory is
not infrared free and we define:
\beq
\mrt = &\mx(0,T) \nonumber \\
\lrt = &\lb(0,T).
\label{threetwentynine}  \eeq
\par
{\bf 3d) The phase transition:}
We have now established the connection between the renormalized quantities
at zero and non-zero temperature. For any given ``short distance''
theory,
through the solution of the evolution equations,
we can determine the renormalized zero temperature theory
in terms of the location of the minimum $\rhz$ and the renormalized
quartic coupling $\lr$.
We can also obtain
$\rhzt$ and $\lrt$
for non-zero temperatures $T < T_{cr}$.
For $T \geq T_{cr}$ the symmetry is
restored ($\rhzt=0$) and the non-zero temperature theory is described
in terms of $\mrt$ and $\lrt$.
We can, therefore, study the phase transition
in terms of renormalized quantities (at zero and non-zero temperature),
without explicit reference to the ``short distance'' theory.
In fig. 7 we plot $\rhzt/\rhz$, $\lrt$ and
$\mrt/T^2$ as a function of temperature, for
$N=4$ and $\lr=0.1$.
As the temperature increases towards
$T_{cr}$ we observe a continuous transition from the spontaneously broken
to the symmetric phase. This clearly indicates a
{\it second order phase transition}.
The renormalized quartic coupling $\lrt$ remains close to its
zero temperature value $\lr$ for a large range of temperatures and
drops quickly to zero at $T=T_{cr}$.
Recalling our parametrization of the effective average potential in terms of
its
successive $\rho$ derivatives at the minimum, we conclude that, at $T_{cr}$,
the first
non-zero term in the expression for the effective potential
is the $\phi^6$ term (which we have neglected so far in our truncated
solution).
For $T \gg T_{cr}$ the coupling
$\lrt$ quickly grows to approximately its zero temperature
value $\lr$, while $\mrt$ asymptotically becomes
proportional to $T^2$ as $T \rightarrow \infty$.
In the symmetric phase the quartic coupling is positive for all temperatures
and vanishes only for $T \rightarrow \tcr$.
\par
Having presented the general features of the non-zero temperature
theory, we now turn to a more detailed quantitative discussion.
The value of the critical temperature $T_{cr}$ in terms of the
zero temperature quantities has been calculated in the context of
perturbation theory \cite{linde, doljack, weinberg}.
It was found that $T_{cr}$ is given by
$T^2_{cr}= \frac{24}{N+2} \rhz$, independently of $\lr$ in lowest order.
In table 1 we list the quantity $\frac{T^2_{cr}(N+2)}{\rhz}$
for various values of $N$ and $\lr$.
We observe excellent agreement
with perturbation theory for $\lr \rightarrow 0$,
and significant deviations for larger $\lr$.
This is not surprising if one recalls that
perturbation theory breaks down in region of size
$|T-T_{cr}| / T_{cr} = {\cal O}(\lr)$. For small $\lr$ this region shrinks to
zero and the value of $T_{cr}$ can be accurately determined.
It should be pointed out that the critical temperature is a non-universal
quantity which
depends on the evolution in all three (low temperature, intermediate and
high temperature) regions. As a result it reflects underlying degrees of
freedom with effective four- and three-dimensional character.
Another quantity which can be compared with the perturbative
predictions is $\mrt$ in the limit $T \rightarrow \infty$.
In table 2 we list the results of our calculation for the expression
$\left[ \frac{\mrt }{\lrt (N+2) T^2} \right]^{-1}$
for $T^2/\rhz = 10^6$.
The perturbative
result for this quantity is 24, independent of $N$ and $\lr$.
We find again excellent agreement for $\lr \rightarrow 0$.
\par
The most important aspect of our calculation is related to the infrared
behaviour of the theory for $T \rightarrow \tcr$.
The temperature dependence of $\rhzt, \lrt, \mrt$ near $\tcr$
can be
characterized by critical exponents. Following
the notation of statistical mechanics, we parametrize the critical behaviour
of $\rhzt$ and $ \mrt$ as follows:
\beq
\rhzt \propto &(\tcr^2-T^2)^{2 \beta} \nonumber \\
\mrt \propto &(T^2-\tcr^2)^{2 \nu}.
\label{threethirty} \eeq
We also define a critical exponent $\zeta$ for $\lrt$ in the symmetric
regime:
\be
\lrt \propto (T^2- \tcr^2)^{\zeta}.
\label{threethirtyone} \ee
These exponents are plotted as function of the logarithm of
$|T^2-\tcr^2|$ in fig. 8 for $N=4$.
(We should note at this point that, in the limit $T \rightarrow \tcr$,
the above definition coincides with the more conventional one,
which is given in terms of $|T-\tcr|$.)
The temperature dependent exponent $2 \beta$ is defined as the
derivative of $\ln \rhz$ with respect to $\ln |T^2 - \tcr^2|$, and similarly
for the other exponents.
We notice that in the limit $T \rightarrow  \tcr$
the critical exponents approach asymptotic values. These are
independent of $\lr$ and therefore fall into universality classes
determined only by $N$. They are equal to the critical exponents of the
zero temperature three-dimensional theory.
This fact can be understood by recalling that the dynamics in the
critical region (near the phase transition) is solely
determined by the three-dimensional fixed point in the high temperature region,
without any memory of the evolution in the low temperature or intermediate
regions.
In table 3 we list the results of our calculation of $\beta, \nu, \zeta$
for various $N$.
The critical exponents $\beta,\nu$
for the three-dimensional theory have been calculated by several
methods: $\epsilon$ expansion,
summed perturbation series in
the symmetric phase in three dimensions, $1/N$ expansion,
lattice calculations.
For comparison with our results we list in table 3 the most
accurate values of $\beta$ and
$ \nu$ obtained by the previously mentioned methods.
The agreement is not perfect, especially for $N=1$,
since we have used very crude truncations and approximations up to this point.
A consistency check is provided by the
scaling laws which give
$\nu = 2 \beta$
in the limit of zero wave function renormalization.
The above relation is satisfied by our numerical
results to very good accuracy.
\par
The critical behaviour of $\lrt$ is related to the resolution of the
problem of the infrared divergences which cause the breakdown of the
standard perturbative expansion in the limit $T \rightarrow \tcr$
\cite{linde, doljack, weinberg}.
The infrared problem is manifest in the presence of higher order
contributions to the effective potential which contain increasing
powers of $\frac{\lrt T}{K}$, where $K$ is the effective
infrared cutoff of the theory. In the standard
perturbative treatments the evolution of $\lbkt$
is omitted and $\lrt$ is approximated by its zero temperature value
$\lr$. Also the temperature dependent
renormalized mass is taken as the infrared
cutoff. It is obvious then
that these contributions diverge and the perturbative expansion
breaks down near the critical temperature or in the presence of Goldstone
modes.
A similar situation appears for the zero temperature three-dimensional
theory in the critical region \cite{parisi}.
In this case the problem results from an effective expansion
in terms of the quantity
$\frac{u}{[M^2-M_{cr}^2]^{1/2}}$, where
$u$ is the bare three-dimensional
quartic coupling and $[M^2-M_{cr}^2]^{1/2}$
is a measure of the distance from the point where the phase
transition occurs as it is approached from the symmetric phase.
The two situations can be seen to be of
identical nature by simply remembering that the non-zero temperature
four-dimensional
coupling $\lb$ corresponds to an effective three-dimensional
coupling $\lb T$ and
that the effective infrared cutoff in the symmetric phase is
equal to $m_R(T)$.
In the three-dimensional
case the problem has been resolved \cite{parisi} by a reformulation of the
calculation in terms of an effective parameter $\frac{\lx_3}{m}$,
where $\lx_3$ is the renormalized 1-PI four point function
in three dimensions (the renormalized quartic coupling) and
$m$ the renormalized mass (equal to the inverse correlation length).
It has been found
\cite{parisi,zinn} that the above quantity has an infrared stable fixed
point in the critical region $m \rightarrow 0.$ No infrared divergences
arise within this approach for the symmetric phase.
(The phase with spontaneous symmetry breaking remains inaccessible for
$N > 1$, except for universal properties at the phase
transition.) Their only residual effect is detected in the
strong renormalization of $\lx_3$.
In our scheme the problem is formulated in terms of the effective
dimensionless parameters
$\kx(k,T) = \frac{\rhzkt}{k T}, \lx(k,T) = \lbkt \frac{T}{k}$
(see equations (\ref{threetwentyfour})),
for which a fixed point corresponding to the phase transition is found.
The critical behaviour is determined by this fixed point in the limit
$k \rightarrow 0$.
Everything remains finite in the vicinity of the critical temperature,
and the only memory of the infrared divergences is reflected in the
strong renormalization of $\lrt$ near $\tcr$.
We conclude that the infrared problem
disappears if formulated in terms of the appropriate
renormalized quantities (see also \cite{freire}).
When expressed in the correct language, it
becomes simply a manifestation of the strong renormalization
effects in the critical region.
In order to compare with the three-dimensional results we have
calculated the quantity $\frac{\lrt T}{m_R(T)}$
in the limit $T \rightarrow \tcr$. We find that it reaches
an asymptotic value depending on $N$, which we list in table 4 for
various $N$. For comparison we quote the results for the infrared
fixed point of $\frac{\lx_3}{m}$
as summarized in \cite{parisi,zinn} and as can be calculated
in the large $N$ limit with other methods \cite{largen}.
Good agreement is observed.
Moreover, the existence of the asymptotic value for
$\frac{\lrt T}{m_R(T)}$
explains the equality of the critical exponents $\nu$ and $\zeta$ which is
apparent in table 3.
\par
{\bf 3e) Improved truncations and wave function renormalization:}
It is apparent from table 3 that the approximations we have used
up till now are too crude for the precise determination of the universal
critical behaviour as this is parametrized by the critical exponents.
We must, therefore, consider improved truncations which capture
more of the relevant dynamics.
We shall not discuss non-universal quantities (such as the critical
temperature or the high temperature mass term) any more. The agreement with
standard perturbation theory is expected to persist
in the cases where perturbation theory is applicable (small $\lr$).
We concentrate on the purely {\it three-dimensional behaviour} as this is
determined by the fixed point. We improve out calculation by
taking into account higher $\rho$-derivatives of the effective average
potential. The third derivative is expected to
be important, since the $\phi^6$ term is a ``relevant'' term for the
three-dimensional theory (actually it is the last ``relevant'' term, often
characterized as ``marginal''). Higher derivatives should induce only small
changes in the evolution near the fixed point, since they correspond to
``irrelevent'' terms. The wave function renormalization is also an important
ingredient, even though the anomalous dimension $\eta$ is known to be
a small quantity for the three-dimensional theory.
We shall not get into the details of the calculation here, but we
give a brief description of the main results.
The reader is referred to ref. \cite{indices} for a thorough presentation.
We use dimensionless three-dimensional couplings according to
\beq
\kx =~&Z_k k^{-1} \rhz
\nonumber \\
\mt =~&Z_k^{-1} k^{-2} \mb = k^{-2} \mx
\nonumber \\
\lx =~&Z_k^{-2} k^{-1} \lb
\nonumber \\
u_n =~&Z_k^{-n} k^{n-3} U_n,
\label{threethirtytwo} \eeq
where we have used the definition (\ref{twotwelve}) and also defined
$U'_k(0)=\mb(k)$,
$U''_k(\rhz)=\lb(k)$ and $U^{(n)}_k(\rhz)=U_{n}(k)$.
(The primes denote $\rho$-derivatives.)
The above relations are the generalization of eqs. (\ref{threetwentyfour}).
The anomalous dimension is given by
\be
\eta(k) = -\frac{d}{dt} (\ln Z_k).
\label{threethirtythree} \ee
We subsequently derive evolution equations for the dimensionless couplings.
We encounter again the necessity of a truncation in order to be able to
solve the infinite system of coupled differential equations.
We use various truncations in order to study the effect of higher derivatives
on the fixed point.
In table 5 the fixed point values of various parameters are listed, for
$N=3$ and various
truncations of the evolution equations.
For the first row only two evolution equations are considered
for $\kx$ and $\lx$, and the higher derivatives of the potential
as well as the anomalous dimension are set to
zero. Then, the number of evolution equations is enlarged to three with the
inclusion of $u_3$. Subsequently $\eta$ and $u_4$ are added.
Finally $u_5$ and $u_6$ are approximated
(their contributions are proportional to $\eta$ which, for the
three-dimensional theory, is $\eta \simeq 0-0.04)$
and are included in the system of five
coupled evolution equations.
It becomes apparent from table 5, that the influence of the higher
derivatives of the potential on the fixed point,
and therefore on the critical dynamics, is small.
In fact, for an efficient truncation it would suffice to consider the
``relevant'' first three $\rho$-derivatives of the potential and the anomalous
dimension. The inclusion of higher derivatives gives small improvements,
in agreement with their characterization as ``irrelevant''.
\par
A typical example of the numerical integration of the evolution
equations is displayed in fig. 9, for $N=3$.
The parameters $u_4$, $u_5$, $u_6$ are used even though they are not plotted.
The evolution starts at $k=\Lambda$ with the classical action, so that
$Z_{\Lambda}=1$, $u_3(\Lambda)=u_4(\Lambda)=u_5(\Lambda)=u_6(\Lambda)=0$.
The quartic coupling is arbitrarily chosen
$\lx(\Lambda)=\lb(\Lambda)/\Lambda=0.1$. Two values of $\kx(\Lambda)$
are selected so that the system is very close to the two
sides of the critical line.
The two values are indistinguishable in the graph
and are given by
$\kx(\Lambda) = \kx_{cr} + \dkl$, with
$\kx_{cr} = [\rho_0(\Lambda)]_{cr}/\Lambda \simeq 0.107$,
$|\dkl| \ll \kx_{cr}$ and $\dkl$ positive or negative.
With these initial conditions the system evolves towards the fixed point in
the critical region and remains close to it for a long ``time''
$|t|=-\ln(k/\Lambda)$.
By staying near the fixed point for
several orders of magnitude in $t$, the system loses memory of the initial
conditions at the cutoff. As a result the critical behaviour is
indepenent of the detailed short distance physics and is
determined by the fixed point, as long as the evolution
starts sufficiently close to the critical line.
In the final part of the evolution, as $k \rightarrow 0$, the theory
deviates from the fixed poin
either towards the phase with spontaneous symmetry breaking
or the symmetric one,
for $\dkl$ positive or negative respectively.
\par
The critical exponents can be determined through the renormalized
expectation values, masses and couplings, which are obtained
in the limit $k \rightarrow 0$.
For the anomalous dimension the relevant value is the fixed
point value $\eta_*$.
The detailed discussion is
presented in ref. \cite{indices}.
Here we only summarize in table 6 the results for various
$N$.
The critical exponents satisfy the scaling laws
\beq
\beta =~&(1+ \eta_*) \frac{\nu}{2}
\label{threethirtyfour} \\
\gamma =~&(2-\eta_*) \nu
\label{threethirtyfive} \eeq
with an accuracy of 0.1\%.
The exponent $\delta$ can also be determined and it satisfies the
well known scaling law
\be
\delta = \frac{5 -\eta_*}{1 +\eta_*}.
\label{threethirtysix} \ee
In table 6 we have included values for the critical exponents obtained
through other methods (summed perturbation series in three dimensions,
$\ex$ expansion, lattice calculations and $1/N$ expansion)
for comparison.
We observe agreement at the 1-5 \% level for the exponents
$\beta$, $\nu$, and $\gamma$. For the anomalous dimension $\eta_*$
we observe satisfactory
agreement of our results with the quoted values, even though
$\eta_*$ is a small quantity and
most severely affected by the residual approximations
in our method.
\par
This concludes the discussion of the high temperature phase transition
for the $O(N)$-symmetric $\phi^4$ theory through the method of
the effective average action.
We have investigated all its aspects concerning the non-universal and
universal behaviour, with full control over the infrared regime.
Moreover, we have obtained accurate
results for such non-trivial quantities as the critical
exponents. In the next section we use alternative methods
in order to study the phase transition in the large $N$ limit.
Apart from providing a check of the conclusions of this section, this
will enable us to make contact with the standard perturbative methods.

\setcounter{equation}{0}
\renewcommand{\theequation}{{\bf 4.}\arabic{equation}}

\section*{4. The high temperature phase transition
in the large $N$ limit}

In the large $N$ limit \cite{largenn} two significant simplifications
appear in the formalism: Firstly, the anomalous dimension $\eta$ vanishes
which allows us to set $Z_k=1$. Secondly, the dynamics is dominated
by the Goldstone modes (in the spontaneously broken phase).
As a result, the analytical study of the evolution equations is
possible and compact expressions for the evolution of the
whole effective average potential can be obtained. These results
are presented in ref. \cite{indices} and verify the analysis of the
previous section. Here, we are interested in studying the high
temperature phase transition with alternative methods. As such we shall use
the saddle point evaluation of the functional integral for the
effective potential in the limit
$N \rightarrow \infty$, and the Schwinger-Dyson
equations in the same limit.
\par
{\bf 4a) The effective potential in the large $N$ limit:}
The first method results in the well known
expression for the effective potential
\cite{amit, zinn, largen, jain}
\be
U'(\rho) = -\mbu + \lb \rho + \frac{N}{2} \lb
\int_{\Lambda} \frac{d^4 q}{(2 \pi)^4} \frac{1}{ q^2 + U'(\rho)},
\label{fourone} \ee
where
$\mbu$, $\lb$ denote bare quantities and we remind the reader that
$\rho = \frac{1}{2} \phi^a \phi_a$ and the derivatives of
$U$ are taken with respect to $\rho$. An ultraviolet
cutoff $\Lambda$ is implied for the momentum integration.
The generalization of the above expression
for non-zero temperature follows
the discussion is subsection 3b and reads
\be
U'(\rho,T) = -\mbu + \lb \rho + \frac{N}{2} \lb
T \sum_m \int_{\Lambda} \frac{d^{3} \vec{q}}{(2 \pi)^{3}}
\frac{1}{ \vec{q}^2 + 4 \pi^2 m^2 T^2 + U'(\rho,T)}.
\label{fourtwo} \ee
\par
It is convenient to define the
functions
\beq
I(T) = &T \sum_m
\int_{\Lambda} \frac{d^3 \vec{q}}{(2 \pi)^3}
\frac{1}{\vec{q}^2+4 \pi^2 m^2 T^2}
\label{fourthree} \\
I_n(w,T) = &T \sum_m
\int_{\Lambda} \frac{d^3 \vec{q}}{(2 \pi)^3}
\frac{1}{(\vec{q}^2+4 \pi^2 m^2 T^2 + w)^n}.
\label{fourfour} \eeq
Then eq. (\ref{fourtwo}) can be written as
\be
U'(\rho,T) = \left[ \frac{N}{2} \lb I(T)-\mbu \right]
+ \lb \rho + \frac{N}{2} \lb
\left[ I_1(U'(\rho,T),T) - I(T) \right].
\label{fourfive} \ee
We are interested in solving the above equation
near the phase transition. This correponds to a temperature region in
which typically $U'(\rho,T) \ll T^2$ near the minimum of the potential.
We can, therefore, use a high temperature expansion in terms of $w/T^2$
for the evaluation of $I_n(w,T)$ in this regime.
In the opposite limit of zero temperature the discrete summation
can be replaced by an integration over a continuous $q^0$
and the integrals become four-dimensional.
Integrals similar to $I(T), I_n(w,T)$ have been discussed extensively in
the literature \cite{doljack, kapusta}. We briefly review here the
results which are relevant for our investigation.
$I(T)$ is given by
\be
I(T) = \frac{\Lambda^2}{16 \pi^2} + \frac{T^2}{12}.
\label{foursix} \ee
The first term
corresponds to the quadratic divergence of the zero temperature theory,
while for the evaluation of the second term we have made use of
$\Lambda \gg T$.
Notice that $I(T)$ receives contributions from
all values of $m$ in the infinite sum. In this sense
it incorporates effects of the effective three-dimensional theory
(modes with $m=0$) as well as the four-dimensional one ($m \not= 0$).
For the difference $I_1(w,T)-I(T)$ we find \cite{doljack}
\be
I_1(w,T)-I(T) = - \frac{1}{4 \pi} T \sqrt{w}
+ \frac{w}{8 \pi^2} \ln \left( \frac{\ct T}{\Lambda} \right),
\label{fourseven} \ee
with $\ct =\exp(\frac{1}{2} + \ln(4\pi) - \gamma) \simeq 11.6$.
The above expression gives the first two terms of the
high temperature expansion.
The leading term $\sim T \sqrt{w}$
comes entirely from the $m=0$ term in the infinite sum.
As a result eq. (\ref{fourseven}) in the leading order for small
$w/T^2$ reflects only
the dynamics of the effective three-dimensional
theory.
In contrast, the logarithmic term is typical for the four-dimensional
renormalization effects from the momentum range between
$\ct T$ and $\Lambda$.
\par
Combining eqs. (\ref{fourfive}), (\ref{foursix}) and keeping only the
leading contribution in (\ref{fourseven})
we obtain for $T^2 \gg U'(\rho,T)$
\be
U'(\rho,T) =
N \lb \frac{T^2}{24}
- \left[ \mbu - N \lb \frac{\Lambda^2}{32 \pi^2} \right]
+ \lb \rho - N \lb \frac{T}{8 \pi} \sqrt{U'(\rho,T)}.
\label{foureight} \ee
The ultraviolet divergences
can be absorbed in the $(T=0)$ renormalization of
the mass parameter and coupling.
We define the renormalized parameters
for the zero temperature four-dimensional
theory \cite{largenn}
\beq
\frac{\mu^2_R}{\lambda_R} = &\frac{\mbu}{\lb} - N \frac{\Lambda^2}{32 \pi^2},
\label{fournine} \\
\lambda_R = &\frac{\lb}{1 + \frac{N \lb}{32 \pi^2} \ln \left(
\frac{\Lambda^2}{M^2} \right) },
\label{fourten} \eeq
where $M$ is an appropriate renormalization scale $M \ll \Lambda$.
Eq. (\ref{foureight}) now reads
\be
U'(\rho,T) = \lb \left[ N \frac{T^2}{24} - \frac{\mu^2_R}{\lambda_R} \right]
+ \lb \rho - N \lb \frac{T}{8 \pi} \sqrt{U'(\rho,T)}.
\label{foureleven} \ee
The critical temperature $T_{cr}$ is defined
as the temperature at which $U'(0,\tcr)=0$.
This gives the well known result
\cite{linde, doljack, weinberg}
\be
T^2_{cr} = \frac{24}{N} \frac{\mu^2_R}{\lambda_R}.
\label{fourtwelve} \ee
We concentrate on temperatures $T \sim \tcr$ and $\rho \ll T^2$, for which
we obtain from eq. (\ref{foureleven}) in leading order
(neglecting $U'(\rho,T)$ on the l.h.s)
\be
U(\rho,T) = \frac{\pi^2}{9} \left( \frac{T^2-\tcr^2}{T} \right)^2 \rho
+ \frac{1}{N} \frac{8 \pi^2}{3} \frac{T^2-\tcr^2}{T^2} \rho^2
+ \frac{1}{N^2} \frac{64 \pi^2}{3} \frac{1}{T^2} \rho^3.
\label{fourthirteen} \ee
We observe the characteristic
$N$ dependence for the renormalized couplings
\cite{largenn}.
Notice that the only surviving reference to the parameters of the
four-dimensional theory
exists in the definition of the critical temperature (\ref{fourtwelve}).
According to our discussion of the integrals
$I, I_1$, the critical temperature is determined by the dynamics of both the
four-dimensional and three-dimensional theory. On the contrary, the
form of the potential in the critical region
(and therefore the critical behaviour) reflects only the
effective three-dimensional theory. This is in complete agreement with
the conclusions of section 3.
\par
Eq. (\ref{fourthirteen}) predicts a {\it second order phase transition}.
For $T \geq \tcr$ the theory is in the symmetric phase with the minimum of the
potential at
$\rho_0 = 0$. As the critical temperature is approached from above
the mass term is given by
\be
m_R(T) \equiv \sqrt{U'(0,T)} = \frac{\pi}{3} \frac{T^2-\tcr^2}{T},
\label{fourfourteen} \ee
a result first obtained in \cite{doljack}.
In the language of critical exponents, the behaviour of $m_R$
corresponds to an exponent
\be
\nu = \lim_{T \rightarrow \tcr}
\frac{d \left[ \ln m_R(T) \right] }{d \left[ \ln(T-\tcr) \right] } = 1.
\label{fourfifteen} \ee
The quartic coupling in the symmetric phase
is given by
\be
\lambda_R(T) \equiv U''(0,T)
= \frac{1}{N} \frac{16 \pi^2}{3} \frac{T^2-\tcr^2}{T^2}.
\label{foursixteen} \ee
It goes to zero as the critical temperature is approached
with an exponent
\be
\zeta = \lim_{T \rightarrow \tcr} \frac{d \left[ \ln \lambda_R(T)
\right] }{d \left[ \ln(T-\tcr) \right] }
= 1.
\label{fourseventeen} \ee
The dimensionless ratio $\frac{\lambda_R(T) T}{m_R(T)}$
approaches a temperature independent value at $\tcr$
\be
\lim_{T \rightarrow \tcr}
\frac{\lambda_R(T) T}{m_R(T)} = \frac{16 \pi}{N}.
\label{foureighteen} \ee
We remind the reader again
that the ratio $\frac{\lb T}{m_R}$ diverges near $\tcr$ and this
is the reason for the breakdown of the standard perturbative
calculations. We see how this divergence disappears when the
bare quartic coupling is replaced by the renormalized one.
At $\tcr$ the $\rho^3$ ($\phi^6$) term is the first non-zero one,
with
a coefficient given in (\ref{fourthirteen}).
For $T < \tcr$ the potential has a minimum at a non-zero value
\be
\rho_0(T) = \frac{N}{24} (\tcr^2-T^2),
\label{fournineteen} \ee
and assumes the form
\be
U(\rho,T) = \frac{1}{N^2} \frac{64 \pi^2}{3} \frac{1}{T^2}
\left( \rho - \rho_0(T) \right)^3,
\label{fourtwenty} \ee
after the addition of a $\rho$ independent constant.
We point out that for $T < \tcr$ the form of the potential
given by eq. (\ref{fourtwenty}) is valid only for $\rho \geq \rhz (T)$
(see eq. (\ref{foureleven})).
In the ``inner'' region $\rho < \rhz (T)$ the potential
is flat (see ref. \cite{convex} for a discussion of the convexity of the
effective potential).
Notice that the minimum of the potential (the order parameter for the
phase transition)
approaches continuously zero as the critical temperature is
approached from below (justifying our conclusion
for a phase transition of the second kind).
Expression (\ref{fournineteen}) predicts a critical
exponent
\be
\beta = \lim_{T \rightarrow \tcr}
\frac{d \left[ \ln \sqrt{\rho_0(T)} \right] }{d \left[ \ln(\tcr-T) \right] }
= 0.5.
\label{fourtwentyone} \ee
Another interesting point is the zero value of the second $\rho$-derivative
of the potential at the minimum for $T< \tcr$. This reflects the fact that
the $\phi^4$ theory is infrared free in the spontaneously broken phase
due to the Goldstone modes.
At non-zero temperature the dependence of the running coupling on the
infrared cutoff $k$ is linear \cite{transition}
(compared to the logarithmic running at zero
temperature). This results in a zero renormalized coupling
$\lambda_R(T) = U''(\rho_0,T)$
in the limit $k \rightarrow 0$.
We see how this physical effect is naturally incorporated in
our result (\ref{fourtwenty}).
\par
All the results derived above are in full qualitative and quantitative
agreement with the results of section 3, as can be easily verified by
inspection of tables 3, 4 and 6.
\par
{\bf 4b) The Schwinger-Dyson equations:}
Another check can be obtained through the study of the
Schwinger-Dyson equations for the $O(N)$-symmetric
$\phi^4$ theory in the symmetric
phase for large $N$ \cite{schwinger, itzykson}.
In fig. 10 we give a graphical representation
of the first three equations.
Only the leading terms in $1/N$ have been included.
The left-hand-side of the equations involves the first three $\rho$-derivatives
of the potential at zero external momentum (the renormalized mass and
four and six point vertices).
Simple vertices denote bare quantities, and lines
with a circle full propagators.
The representation is schematic and the signs and combinatoric factors for the
various terms have not been included.
The solution of these equations can be carried out along the lines of the
investigation in subsection 4a. We find \cite{largen}
\beq
U'(0,T) = &m^2_R(T) = \left[ \frac{\pi}{3} \frac{T^2-\tcr^2}{T} \right]^2
\nonumber \\
U''(0,T) = &\lr(T) = \frac{1}{N} \frac{16 \pi^2}{3} \frac{T^2-\tcr^2}{T^2}
\nonumber \\
U'''(0,T) = &\frac{1}{N^2} 128 \pi^2 \frac{1}{T^2},
\label{fourtwentytwo} \eeq
in exact agreement with subsection 4a.
The above solutions are again independent of the parameters of the
bare theory. Their universal behaviour is a result of the strong
infrared singularities of the loop diagrams on the r.h.s. of the
equations in fig. 10, in the limit $T \rightarrow \tcr$,
$m_R(T) \rightarrow 0$.
\par
The most important intuition gained from the study of
the Schwinger-Dyson equations concerns the perturbative expansion.
It is well known \cite{doljack}
that the Schwinger-Dyson equation for the propagator (``gap equation'')
for the $\phi^4$ theory in the large $N$ limit (eq. (a) in fig. 10),
can be reproduced
by the summation of an infinite class of diagrams, usually characterized as
``daisy'' or ``cactus'' graphs \cite{doljack,linde}.
The solution of this equation gives the value of the critical temperature
and the temperature dependence of the mass term in the critical
region
\footnote{For a study of high temperature phase transitions for
gauge theories through ``gap equations'' see ref. \cite{buch}.}.
In a similar way, eq. (b) of fig.10 can be turned into a sum
of perturbative contributions involving high powers of the bare coupling
through an iterative solution (i.e. truncation at a finite power of
$\lb$ and successive substitutions on the r.h.s.).
Usually the terms with large powers of $\lb$ can
be neglected for sufficiently small $\lb$. It is clear that this is not
the case near the critical temperature, where
infrared divergences appear at higher loops.
However, the summation of an infinite number
of perturbative contributions as given by the
Scwinger-Dyson equation results in the finite answer of eq.
(\ref{fourtwentytwo}). Similar remarks apply to the $\phi^6$ coupling.
It should also be noted that, if the renormalization of the couplings is
not properly taken into account, perturbative calculations for the
effective potential may lead to predictions for a weakly
first order phase
transition for the scalar theory \cite{wrong}.
A more careful treatment, however, indicates that the barrier in the
perturbative effective potential appears in a region where the
perturbative expansion does not converge due to the infrared divergences
\cite{zwirner}.

\setcounter{equation}{0}
\renewcommand{\theequation}{{\bf 5.}\arabic{equation}}

\section*{5. Other applications of the effective average action and
conclusions}

In this final section we briefly summarize other applications of the
formalism of the effective average action, work in progress and
prospects for future studies.
\par
Work in the past has described correctly the phase structure of the
two- and three-dimensional scalar theories (including the Kosterlitz-Thouless
phase transition) \cite{christof2}.
The discussion of the high temperature phase transition for the
four-dimensional theory, as it was reviewed
in the main body of this work, covered most of its qualitative and
quantitative aspects. The improvement on the accuracy for the critical
exponents and the study of the equation of state for the critical
theory are two directions for further work. The next step is the study of
more complicated scalar theories. Work in progress focuses on
the zero and high temperature phase structure of
a two scalar model with a $Z_2 \times Z_2$ symmetry \cite{preps}.
\par
The idea of averaging as a means to integrate out
field fluctuations
was extended to include fermions
in ref. \cite{fermion}.
\par
Another question which has been answered in the context of the
effective average action concerns the convexity of the effective potential.
In ref. \cite{convex} it was shown that the effective average potential
becomes flat in the inner region (between the two minima) when
quantum fluctuations are integrated out.
The convex effective potential is approached as $k \rightarrow 0$
with a behaviour at the origin given by
\be
\lim_{k \rightarrow 0} U_k'(0) \sim -k^2.
\label{fiveone} \ee
\par
{}From its construction (integration of flucuations with characteristic
momenta larger than a given infrared cutoff $k$),
the effective average action is the appropriate quantity for the study of
physics at a scale $k$.
In practice one has to identify
$k$ with the appropriate physical infrared cutoff
scale of the problem
(such as the Hubble parameter for inflationary cosmology).
In the case of a first order phase transition, one usually wishes to
integrate out the large momentum fluctuations in order to study
physical quantities such as tunneling rates etc. in the resulting effective
potential. It is clear that the classical potential is not in general
a good approximation for such studies. This is more obvious in the cases where
a first order phase transition is the result of radiative corrections
\cite{colwein}.
However, the absence of non-convex parts in the effective potential
(which results from the incorporation in it of the fluctuations associated with
tunneling) makes it an inappropriate quantity too.
The effective average action
(the equivalent of the ``coarse grained effective action''
of statistical mechanics)
can be used with an appropriate choice of $k$. As an example,
the study of critical bubbles during a high temperature first order
phase transition can be carried out using the non-convex effective average
potential, with the scale $k$ determined by a quantity such as
the correlation length in any of the two minima, or the bubble radius or
thickness (a detailed investigation is necessary for specific models).
\par
Finally, the concept of the effective integration of degrees of freedom through
averaging has been developed for gauge fields
\cite{gauge}, and an exact evolution equation has been
formulated for gauge theories \cite{prepg}. Study of this evolution equation
predicts a linear evolution with $k$ for the effective three-dimensional
gauge coupling $e^2 T$. In the case of the Abelian Higgs model this running
results in a smaller coupling as $k$ decreases; whereas, in the
case of the non-Abelian Higgs model (due to the opposite sign in the
$\beta$-function for the gauge coupling) the coupling increases in the
infrared and ``three-dimensional confinement'' appears near the phase
transition and in the symmetric phase \cite{gauge}.
More work in this direction
will ultimately lead to a reliable description and
understanding of the details of the electroweak phase transition.

\newpage
\section*{Figures}

\renewcommand{\labelenumi}{Fig. \arabic{enumi}}
\begin{enumerate}
\item  %Fig. 1
Graphical representation of the exact evolution equation for the
effective average action $\Gamma_k$.
\vspace{6mm}
\item  %Fig. 2
The integrals $L^4_1(w), L^4_2(w)$ for $b=3$ and $a = 0.952$.
\vspace{6mm}
\item  %Fig. 3
$L^4_1(w,T)/L^4_1(w)$ as a function of $T/k$,
for various values of $w/k^2$ ($b=3$, $a = 0.952$).
\vspace{6mm}
\item   %Fig. 4
The evolution of
$\rhz$ at various temperatures.
For $T>T_{cr}$ the evolution of the mass term in the symmetric regime
is also displayed. $N=4$, $\lr=0.1$ ($b=3$, $a=0.952$).
\vspace{6mm}
\item   %Fig. 5
The evolution of $\lb$ at various temperatures. $N=4$, $\lr=0.1$
($b=3$, $a=0.952$).
\vspace{6mm}
\item   %Fig. 6
The phase diagram of the three-dimensional theory for $N=4$,
in a truncation which neglects derivatives of the
potential higher than the second ($b=3$, $a=0.952$).
\vspace{6mm}
\item   %Fig. 7
$\rhzt, \lrt, \mrt$ for a wide range of temperatures. $N=4, \lr=0.1$
$(b=3, a=0.952)$.
\vspace{6mm}
\item   %Fig. 8
The critical exponents as $T_{cr}$ is approached. For $T \rightarrow T_{cr}$
they become equal to the critical exponents of the zero temperature
three-dimensional theory. $N=4$ $(b=3, a=0.952$).
\vspace{6mm}
\item  %Fig. 9
Solution of the evolution equations for $\kx(k)$, $\lx(k)$, $u_3(k)$ and
$\eta(k)$ in the critical region, for initial values slightly above and below
the critical line.
The fixed point for the second order phase transition is displayed, as well as
the final running towards
the phase with spontaneous symmetry breaking or the
symmetric one. $N =3$ ($b=1, a=0.5$).
\vspace{6mm}
\item  %Fig. 10
Schwinger-Dyson equations in the large $N$ limit.

\end{enumerate}

\newpage

\section*{Tables}

\begin{table} [h]
\renewcommand{\arraystretch}{1.5}
\hspace*{\fill}
\begin{tabular}{|c||c|c|c|}     \hline
  $\hspace{1cm}$
 &{$\lr=0.01$} & $\lr=0.1$ & $\lr=1$
\\ \hline \hline
$N=1$ & 24.05 & 24.50 & 26.82
\\ \hline
$N=3$ & 24.03 & 24.48 & 26.70
\\ \hline
$N=4$ & 24.03 & 24.48 & 26.61
 \\ \hline
$N=10$ & 24.02 & 24.42 & 26.10
\\ \hline
\end{tabular}
\hspace*{\fill}
\renewcommand{\arraystretch}{1}
\caption[y]
{$\frac{T_{cr}^2 (N+2)}{\rhz}$ for various $\lr$ and $N$. $b=3, a=0.952$.
Standard perturbation theory gives $\frac{T_{cr}^2 (N+2)}{\rhz}= 24$.}
\end{table}

\renewcommand{\arraystretch}{1.5}
\hspace*{\fill}
\begin{tabular}{|c||c|c|c|}     \hline
  $\hspace{1cm}$
 &{$\lr=10^{-4}$} & $\lr=0.01$ & $\lr=0.1$
\\ \hline \hline
$N=1$ & 23.92 & 23.62 & 23.19
\\ \hline
$N=3$ & 23.94 & 23.91 & 24.02
\\ \hline
$N=4$ & 23.95 & 24.01 & 24.32
 \\ \hline
$N=10$ & 23.99 & 24.40 & 25.53
\\ \hline
\end{tabular}
\hspace*{\fill}
\renewcommand{\arraystretch}{1}
\caption[y]
{$\left[ \frac{\mrt }{\lrt (N+2) T^2} \right]^{-1}$ for
$\frac{T^2}{\rhz}=10^6$ and various $\lr$, $N$. $b=3, a=0.952$.
Standard perturbation theory gives $\left[ \frac{\mrt }{\lrt (N+2) T^2}
\right]^{-1} = 24 $ for $\frac{T^2}{\rhz} \rightarrow \infty $.}
\end{table}

\begin{table} [h]
\renewcommand{\arraystretch}{1.5}
\hspace*{\fill}
\begin{tabular}{|c|||c|c|||c|c|||c|c|}  \hline
  $\hspace{1cm}$
 &\multicolumn{2}{c|||}{$\beta$}
&\multicolumn{2}{c|||}{$\nu$} & \multicolumn{2}{c|}{$\zeta$}
\\
\hline \hline
$N=1$ & 0.25 & $0.33^{(a)}~ 0.31^{(b)}
$ & 0.50 & $0.63^{(a,b)}$ & 0.50 & $=\nu^{(d)}$
\\ \hline
$N=3$ & 0.37 & $0.37^{(a)}~ 0.38^{(b)}$
& 0.75 & $0.71^{(a,b)}$ & 0.75 & $=\nu^{(d)}$
\\ \hline
$N=4$ & 0.40 & & 0.81 & & 0.81 & $=\nu^{(d)}$
\\ \hline
$N=10$ & 0.46 & $0.45^{(c)}$ & 0.92
& $0.88^{(c)}$
& 0.92 & $=\nu^{(d)}$
\\ \hline
$N=\infty$ & & $0.5^{(c)}$ & & $1^{(c)}$ & & $=\nu^{(d)}$
\\ \hline
\end{tabular}
\hspace*{\fill}
\renewcommand{\arraystretch}{1}
\caption[y]
{Critical exponents for various $N$ ($b=3, a=0.952$).
For comparison we list results obtained with other methods as summarized in
\cite{zinn} and \cite{journal}: \\
a) From summed perturbation series in fixed dimension 3 at six-loop order,
or from the $\ex$ expansion at order $\ex^5$. \\
b) From lattice calculations. \\
c) From the $1/N$ expansion at order $1/N^2$. \\
d) From the finite value of $\frac{\lambda_3}{m}$
at the critical point as given by summed perturbation series,
where $\lambda_3$ is the renormalized
three-dimensional coupling
and $m$ the renormalized mass \cite{parisi,zinn}.
}
\end{table}

\begin{table} [h]
\renewcommand{\arraystretch}{1.5}
\hspace*{\fill}
\begin{tabular}{|c||c|c|c|c|}   \hline
{$N$} & 1 & 3 & 4 & 10
\\ \hline
$\frac{\lrt T}{m_R(T)}$ & 6.8 & 5.5 & 4.8 & 3.7
\\ \hline
$\frac{\lambda_3}{m}$ & $7.9^{(a)}$ & $6.4^{(a)}$ & & $5.0^{(b)}$
\\ \hline
\end{tabular}
\hspace*{\fill}
\renewcommand{\arraystretch}{1}
\caption[y]
{The asymptotic value of $\frac{\lrt T}{m_R(T)}$ in the limit
$T \rightarrow T_{cr}$ for various $N$ ($b=3, a=0.952$).
For comparison we have listed the values of $\frac{\lambda_3}{m}$
at the critical point obtained through other methods: \\
a) From summed perturbation series in fixed dimension 3 \cite{parisi,zinn}. \\
b) From large $N$ calculations \cite{largen}.
}
\end{table}

\begin{table} [h]
\renewcommand{\arraystretch}{1.5}
\hspace*{\fill}
\begin{tabular}{|c|c|c|c|c|c|c|}        \hline
$\kx_*$
& $\lx_*$
& $u_{3*}$
& $\eta_*$
& $u_{4*}$
& $u_{5*}$
& $u_{6*}$
\\ \hline \hline
$6.57 \times 10^{-2}$
& 11.5
&
&
&
&
&
\\ \hline
$8.01 \times 10^{-2}$
& 7.27
& 52.8
&
&
&
&
\\ \hline
$7.86 \times 10^{-2}$
& 6.64
& 42.0
& $3.58 \times 10^{-2}$
&
&
&
\\ \hline
$7.75 \times 10^{-2}$
& 6.94
& 43.5
& $3.77 \times 10^{-2}$
& 95.7
&
&
\\ \hline
$7.71 \times 10^{-2}$
& 7.03
& 43.4
& $3.83 \times 10^{-2}$
& 111
& $-1.43 \times 10^3$
& $3.72 \times 10^4$
\\ \hline
\end{tabular}
\hspace*{\fill}
\renewcommand{\arraystretch}{1}
\caption[y]
{
Fixed point values for the rescaled minimum of the
effective average potential, its derivatives and the anomalous
dimension, for various truncations of the infinite system
of evolution equations. $N=3$ ($b=1, a=0.5$).
}
\end{table}

\begin{table} [h]
\renewcommand{\arraystretch}{1.5}
\hspace*{\fill}
\begin{tabular}{|c||c|c||c|c||c|c||c|c|}        \hline
$N$
&\multicolumn{2}{c||}{$\beta$}
&\multicolumn{2}{c||}{$\nu$}
&\multicolumn{2}{c||}{$\gamma$}
&\multicolumn{2}{c|}{$\eta_*$}
\\
\hline \hline

&
&$0.3250(15)^a$
&
&$0.6300(15)^a$
&
&$1.2405(15)^{a}$
&
&$0.032(3)^{a}$
\\
1
&0.333
&$0.3270(15)^{b}$
&0.638
&$0.6310(15)^{b}$
&1.247
&$1.2390(25)^{b}$
&0.045
&$0.0375(25)^{b}$
\\

&
&$0.312(5)^{c}$
&
&$0.6305(15)^{c}$
&
&$1.2385(25)^{c}$
&
&
\\ \hline

&
&$0.3455(20)^{a}$
&
&$0.6695(20)^{a}$
&
&$1.316(25)^{a}$
&
&$0.033(4)^{a}$
\\
2
&0.365
&$0.3485(35)^{b}$
&0.700
&$0.671(5)^{b}$
&1.371
&$1.315(7)^{b}$
&0.042
&$0.040(3)^{b}$
\\

&
&
&
&$0.672(7)^{c}$
&
&$1.33(2)^{c}$
&
&
\\ \hline

&
&$0.3645(25)^{a}$
&
&$0.705(3)^{a}$
&
&$1.386(4)^{a}$
&
&$0.033(4)^{a}$
\\
3
&0.390
&$0.368(4)^{b}$
&0.752
&$0.710(7)^{b}$
&1.474
&$1.39(1)^{b}$
&0.038
&$0.040(3)^{b}$
\\

&
&$0.38(3)^{c}$
&
&$0.715(20)^{c}$
&
&$1.38(2)^{c}$
&
&
\\ \hline
4
&0.409
&
&0.791
&
&1.556
&
&0.034
&
\\  \hline
10
&0.461
&$0.449^{d}$
&0.906
&$0.877^{d}$
&1.795
&$1.732^{d}$
&0.019
&$0.025^{d}$
\\ \hline
20
&0.481
&$0.477^{d}$
&0.952
&$0.942^{d}$
&1.895
&$1.872^{d}$
&0.010
&$0.013^{d}$
\\ \hline
100
&0.497
&$0.496^{d}$
&0.992
&$0.989^{d}$
&1.981
&$1.975^{d}$
&0.002
&$0.003^{d}$
\\ \hline
\end{tabular}
\hspace*{\fill}
\renewcommand{\arraystretch}{1}
\caption[y]
{
Critical exponents of the three-dimensional theory for various values of $N$
($b=1, a=0.5$).
For comparison we list results obtained with other methods as summarized in
\cite{zinn} and \cite{journal}: \\
a) From summed perturbation series in fixed dimension 3 at six-loop order. \\
b) From the $\ex$ expansion at order $\ex^5$. \\
c) From lattice calculations. \\
d) From the $1/N$ expansion at order $1/N^2$.
}
\end{table}

\end{document}